\documentclass[aps,prb,twocolumn,showpacs,superscriptaddress]{revtex4-1}
\usepackage{graphicx}

\begin{document}

\title{Zn-induced spin dynamics in overdoped La$_{2-x}$Sr$_x$Cu$_{1-y}$Zn$_y$O$_4$}

\author{Stephen D. Wilson}
\affiliation{
Department of Physics, Boston College, Chestnut Hill, Massachussetts 02467, USA}
\author{Z. Yamani}
\affiliation{ Chalk River Laboratories, Canadian Neutron Beam Centre, National Research Council, Chalk River, Ontario, Canada K0J 1P0} 
\author{Chetan Dhital}
\affiliation{
Department of Physics, Boston College, Chestnut Hill, Massachussetts 02467, USA}
\author{B. Freelon}
\affiliation{ Physics Department, University of California, Berkeley, California 94720, USA}
\author{P. G. Freeman}
\affiliation{Institut Laue Langevin, 6 Rue Jules Horowitz BP 156, F-38042 Grenoble Cedex 9, France}
\affiliation{Hemlholtz-Zentrum Berlin, Hahn-Mietner-Platz 1, 14109 Berlin, Germany}
\author{J. A. Fernandez-Baca}
\affiliation{ Neutron Scattering Science Division, Oak Ridge National Laboratory, Oak Ridge, Tennessee 37831-6393, USA}
\author{K. Yamada}
\affiliation{ WPI Research Center, Advanced Institute for Materials Research, Tohoku University, Sendai 980-8577, Japan}
\author{S. Wakimoto}
\affiliation{ Quantum Beam Directorate, Japan Atomic Energy Agency, Tokai, Ibaraki 319-1195, Japan}
\author{W. J. L. Buyers}
\affiliation{ Chalk River Laboratories, Canadian Neutron Beam Centre, National Research Council, Chalk River, Ontario, Canada K0J 1P0}
\author{R. J. Birgeneau}
\affiliation{ Physics Department, University of California, Berkeley, California 94720, USA}
\affiliation{ Materials Science Division, Lawrence Berkeley National Lab, Berkeley, Caifornia 94720, USA }
\affiliation{ Materials Science Department, University of California, Berkeley, California 94720, USA}

\begin{abstract}
Spin fluctuations and the local spin susceptibility in isovalently Zn-substituted La$_{2-x}$Sr$_{x}$Cu$_{1-y}$Zn$_y$O$_4$ ($x=0.25$, $y\approx0.01$) are measured via inelastic neutron scattering techniques.  As Zn$^{2+}$ is substituted onto the Cu$^{2+}$-sites, an anomalous enhancement of the local spin susceptibility $\chi^{\prime\prime}(\omega)$ appears due to the emergence of a commensurate antiferromagnetic excitation centered at wave vector \textbf{Q}$=(\pi, \pi, 0)$ that coexists with the known incommensurate SDW excitations at \textbf{Q}$_{HK}=(\pi\pm\delta,\pi), (\pi,\pi\pm\delta)$. Our results support a picture of Zn-induced antiferromagnetic (AF) fluctuations appearing through a local staggered polarization of Cu$^{2+}$-spins, and the simultaneous suppression of T$_c$ as AF fluctuations are slowed in proximity to Zn-impurities suggests the continued importance of high energy AF fluctuations at the far overdoped edge of superconductivity in the cuprates.       
\end{abstract}


\maketitle

\section{Introduction}
Substituting nonmagnetic Zn$^{2+}$ impurities into Cu$^{2+}$-sites within the CuO$_2$ planes of the high temperature superconducting (high-T$_c$) cuprates has long been known to suppress rapidly their superconducting T$_c$\cite{xiao, fukuzumi, xiaoYBCO, tarascon, alloulreview}.  The microscopic mechanism of this suppression and its resulting effect on both the static and dynamic spin behavior of the cuprates remain intensively studied avenues for exploring the electronic interactions within the high-T$_c$ ground state.  While Zn-doping readily destroys superconductivity (SC) in the cuprates at concentrations of only a few percent ($\approx 3\%$)\cite{xiao, xiaoYBCO}, concomitant to this suppression, dramatic modifications are also affected within their residual spin behavior\cite{wakimotoLSCZO, kimura, hirotareview}.  The Zn-induced spin response coincident with the suppression of T$_c$ has motivated a number of studies probing the spin spectrum of the cuprates as a function of nonmagnetic impurity substitution\cite{wakimotoLSCZO, kimura, kimuraoverdoped, adachineutron, matsuda}; many with hopes of identifying spin correlations supporting pair formation as well as exploring competing magnetic ground states emergent as SC vanishes.  

Previous studies have demonstrated that a strong enhancement of local Curie-type magnetism results with increasing Zn substitution into the cuprates suggestive of a competition between SC and local moment magnetism\cite{nakano, alloul, nakano2, mahajan, hanakil}.  More importantly, this enhancement in the local moment magnetism is accompanied by clear changes in the correlated spin response in a variety of Zn-doped cuprate compounds\cite{kimura, wakimotoLSCZO, keimer}. In particular, the introduction of nonmagnetic Zn impurities has strongly suggested the stabilization of microscopically phase separated regions of coexisting, staggered, antiferromagnetic (AF) correlations excluded from the high-T$_c$ superfluid\cite{nachumi, risdiana, julien}.  Unfortunately, due to the complexity of the Zn-induced magnetic response across the phase diagrams of cuprate systems and the potential for interaction with competing phases arising from stripe- or psuedogap-based symmetry breaking, the microscopic details of this phase-separated or ``swiss-cheese'' model\cite{nachumi} of spin behavior remain elusive.       

Doping Zn within the La$_{2-x}$Sr$_{x}$Cu$_{1-y}$Zn$_y$O$_{4}$ (LSCZO(x,y)) matrix yields a complex picture of impurity induced magnetism with a strong dependence on the hole-carrier concentration.  In slightly underdoped LSCZO, Zn-doping is known to shift spectral weight out of the spin gap of LSCO and to stabilize elastic or quasielastic SDW order\cite{hirotareview}.  Contrary to a naive picture of a simple stabilization of stripe magnetic order due to the suppression of superconductivity, Zn-doping in nearly optimally doped LSCO instead appears to induce subgap magnetic states localized about the Zn-impurity sites and correlated between impurities\cite{kimura}.  Slightly overdoped LSCO(x=0.21) samples exhibit similar behavior where Zn impurities stabilize glassy SDW order\cite{kimuraoverdoped} reflecting a direct competition between the stabilized SDW order and the residual SC state.  

Recent experiments however by Wakimoto et al. in far overdoped LSCO(x=0.25) have revealed that Zn-doping fails to stabilize static SDW order and that, instead, the entirety of the resolvable local spin susceptibility of LSCO(x=0.25) is dramatically enhanced while retaining the characteristic energy scale of Zn-free LSCO(x=0.25)\cite{wakimotoLSCZO}.  The purely dynamic response of Zn-doping within this far overdoped regime is potentially reflective of the diminished influence of competing instabilities within the phase diagram of overdoped LSCO\cite{kastner}---such as stripe-order---and suggests that the far overdoped regime at the edge of superconductivity potentially presents a cleaner window for resolving the fundamental nature of Zn-induced magnetic states and their relation to the high-T$_c$ condensate. To-date, however, the origin of the anomalous Zn-induced enhancement in $\chi^{\prime\prime}(\omega)$ within this overdoped LSCO system remains unresolved.

In this article, we present inelastic neutron scattering measurements probing the Zn-induced spin response in far overdoped La$_{1.75}$Sr$_{0.25}$Cu$_{0.99}$Zn$_{0.01}$O$_{4}$.  Our measurements resolve a localized AF mode at \textbf{Q}$_{HK}=(\pi,\pi)$ that constitutes the origin of the anomalous Zn-induced enhancement in the local spin susceptibility of overdoped LSCO \cite{wakimotoLSCO}.  This localized impurity mode results from staggered AF fluctuations induced via the polarization of Cu$^{2+}$ spins in proximity to Zn-impurities within an electronically phase separated ground state; albeit with a characteristic frequency enhanced relative to Zn-induced states previously observed in underdoped LSCZO samples\cite{kimura}.  Our results demonstrate that as LSCZO is overdoped beyond the psuedogap phase within the LSCO phase diagram \cite{taillfeur} Zn-impurities begin to stabilize AF fluctuations reminiscent of the parent system rather than the incommensurate spin excitations that are known to be induced along the stripe-ordering wave vectors in LSCZO concentrations with x$<0.24$ \cite{kimura, kimuraoverdoped}.

\begin{figure}[t]
\includegraphics[scale=.45]{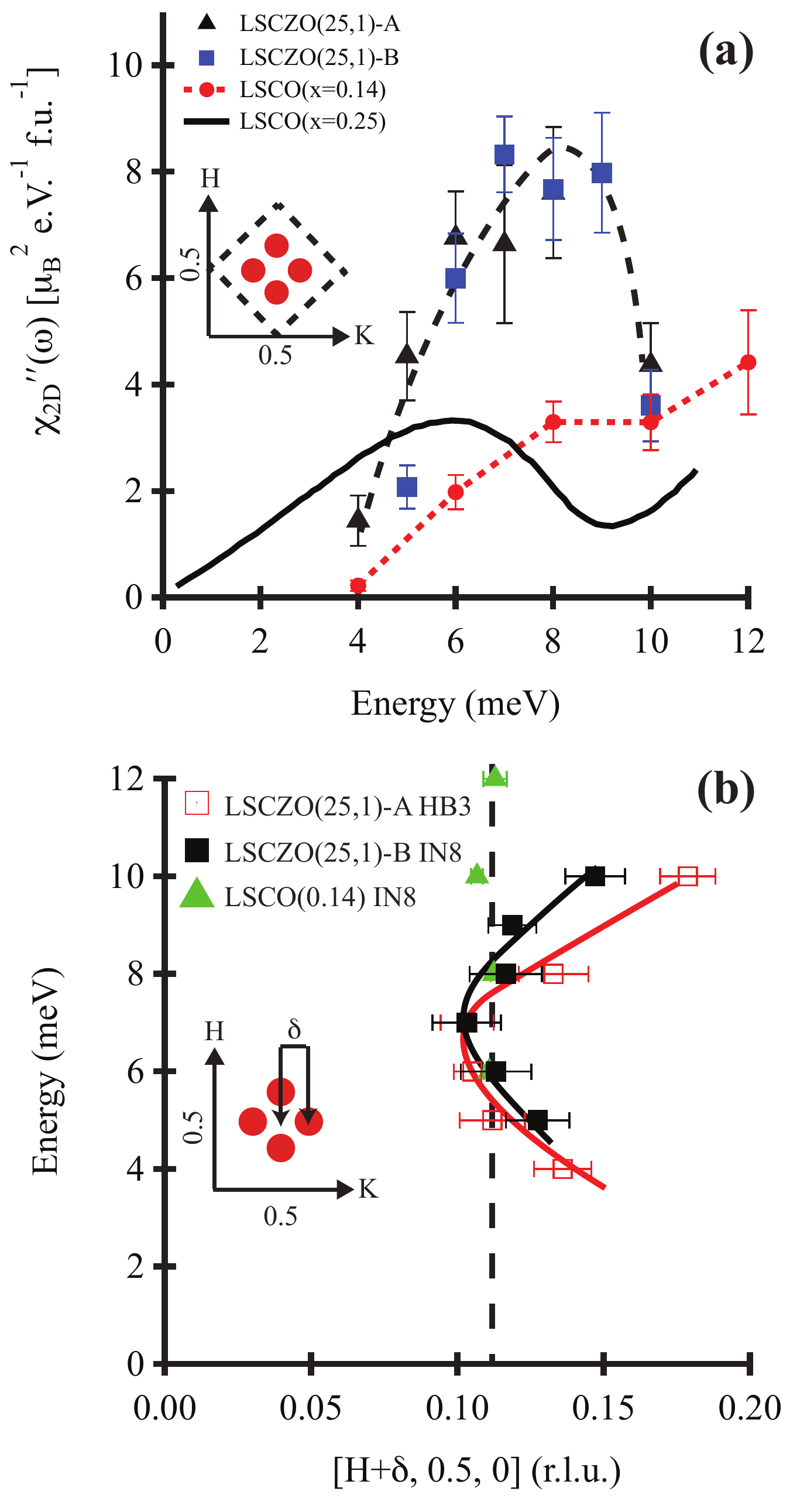}
\caption{(a) $\chi_{2D}^{\prime\prime}(\omega)$ for samples LSCZO(25,1)-B (at T$=5$ K), LSCZO(25,1)-A (at T$=10$ K), and LSCO(x=0.142)(at T$=5$ K). Dashed lines are guides to the eye, and the inset shows the area of the magnetic Brillioun zone integrated.  The solid black line denotes the expected response in $\chi_{2D}^{\prime\prime}(\omega)$ for Zn-free LSCO(x=0.25) from Ref. 23.  (b) Dispersion of IC spin excitations a distance $\delta$ from the (0.5, 0.5 ,0) position for $0\leq E\leq 15$ meV in both LSZO(25,1)-A and LSCZO(25,1)-B samples.  Dashed line highlights the nearly dispersionless IC peak positions in the LSCO(x=0.142) sample at $\delta=0.111\pm0.001$ r.l.u.  The dispersion spectrum was fit with a model of Gaussians symmetrically distributed about (0.5, 0.5, 0).}   
\end{figure}

\section{Experimental details}          
High quality single crystals of La$_{1.75}$Sr$_{0.25}$Cu$_{1-y}$Zn$_{y}$O$_{4}$ (LSCZO(25,y)) were grown via the traveling-solvent floating zone method inside of an infrared mirror furnace as described in earlier work\cite{wakimotoLSCZO}.   Two sets of crystals were grown with nominal starting Zn concentrations of y=0.02 (sample LSCZO(25,1)-A) and y=0.01 (sample LSCZO(25,1)-B) and EDS measurements rendered final Zn concentrations of $y=0.011$ and $y=0.007$ respectively.  Given the large sample volumes of approximately $18$g grown for each sample array, these two Zn concentrations are within error of one another and will be referenced as $y\approx1$ for the remainder of the manuscript. The superconducting T$_c$ onset for the LSCZO(25,1)-A sample was T$_c\approx 6$K and the LSCZO(25,1)-B sample array were the same as those previously characterized in Ref. 5 with an onset T$_c\approx5$K.  Crystals for each sample concentration were coaligned within the [H, K, 0] scattering plane for neutron scattering experiments.  Neutron experiments were performed on the C5 triple-axis spectrometer at the Canadian Neutron Beam Centre at Chalk River Laboratories, the HB-3 triple-axis spectrometer at the High-flux Isotope Reactor (HFIR) at Oak Ridge National Lab, and on the IN8 triple-axis spectrometer at the Institut Laue Langevin (ILL). 

Experiments on C5 were performed with a vertically focusing pyrolitic graphite (PG) monochromator and analyzer using the (0,0,2) reflection, a fixed E$_f=14.5$ meV with one PG filter after the sample, and collimations of $33^\prime-48^\prime-51^\prime-144^\prime$ before the monochromator, sample, analyzer, and detector respectively.  Experiments on HB-3 were performed with vertically focusing PG(002) monochromator and analyzers, fixed E$_f=14.7$ meV with one PG filter after the sample, and collimations of $40^\prime-60^\prime-80^\prime-120^\prime$. Experiments on IN8 were performed using a double focusing Si(111) monochromator comprised of physically bent Si crystals and a double focusing PG(002) analyzer, a fixed E$_f=14.7$ meV, and open collimations. Samples for each experiment were mounted within closed cycle refrigerators.  For the remainder of this paper, positions in reciprocal space are denoted using reciprocal lattice unit notation where \textbf{Q}$(H, K, L)$[r.l.u.]$=(\frac{2\pi}{a}H\vec{h}$, $\frac{2\pi}{b}K\vec{k}$, $\frac{2\pi}{c}L\vec{l})$ [$\AA^{-1}$] using the tetragonal unit cell with $a=b=3.74\AA$, $c\approx13 \AA$ for LSCZO(25,1).  

Local spin susceptibility, which requires integration of  $\chi^{\prime\prime}(Q, \omega)$ over the Brillouin zone, was obtained via line-shape parameters from 1D cuts in reciprocal space through equivalent incommensurate spin density wave vectors. An assumption of isotropic intrinsic line-shapes within the [H,K]-plane due to the tetragonal symmetry was employed, and this assumption was verified via 2D cuts at select energies in each sample.  The resulting $\chi^{\prime\prime}(\omega)$ was converted to [$\mu_B^2$ e.V.$^{-1}$ f.u.$^{-1}$] via normalization to transverse acoustic phonons at the \textbf{Q}=(1.1, 0.9, 0) and \textbf{Q}=(2, 0.2, 0) positions.  While this procedure effectively integrates the spectral weight within the [H, K, 0] plane, the comparatively loose out-of-plane resolution renders this normalization a slight overestimate of $\chi_{2D}^{\prime\prime}(\omega)=\int\int\chi^{\prime\prime}(Q,\omega)\partial q_H \partial q_K $.  The same LSCZO(25,1)-B sample was measured on all three triple-axis spectrometers (C5, HB-3, and IN-8) and our normalization procedure agreed within $20\%$ between experiments.  This overall uncertainty in the absolute moment does not affect relative changes in local susceptibility within the same sample and is not reflected in the experimental error estimates plotted.

\begin{figure}
\includegraphics[scale=.45]{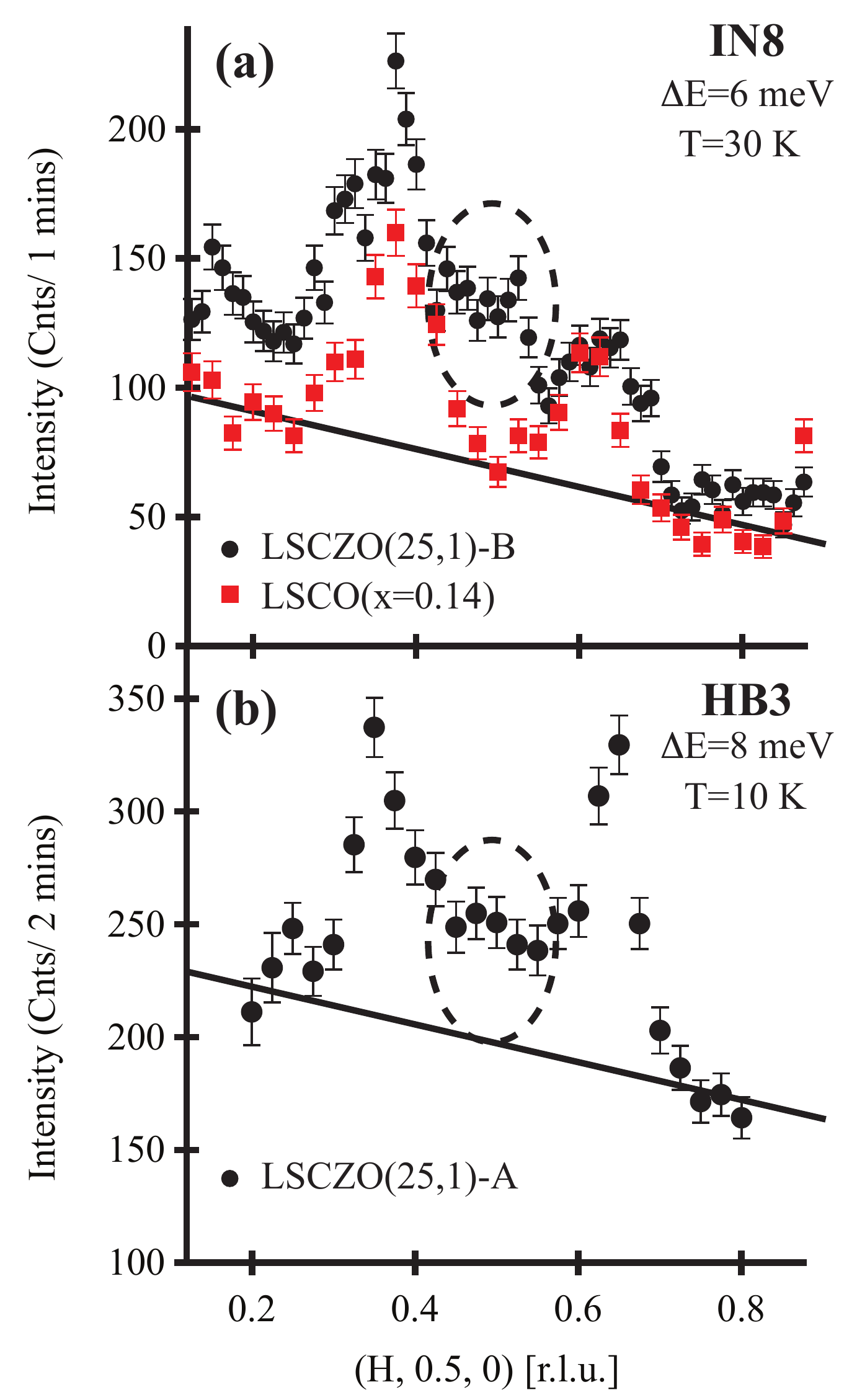}
\caption{Representative constant energy scans taken along the (H, 0.5 ,0) direction in momentum space for the (a) LSCZO(25,1)-B and (b) LSCZO(25,1)-A samples.  Both scans show preliminary hints of spectral weight centered at the (0.5, 0.5, 0) position and distinct from the expected IC peak positions.  A constant energy scan at 6 meV showing the narrower IC peaks of LSCO(x=0.14) with no spectral weight at (0.5, 0.5, 0) is overplotted as red squares in panel (a). Dashed circles highlight the appearance of this anomalous spectral weight at the commensurate AF wave vector.}
\end{figure}

\begin{figure}
\includegraphics[scale=.35]{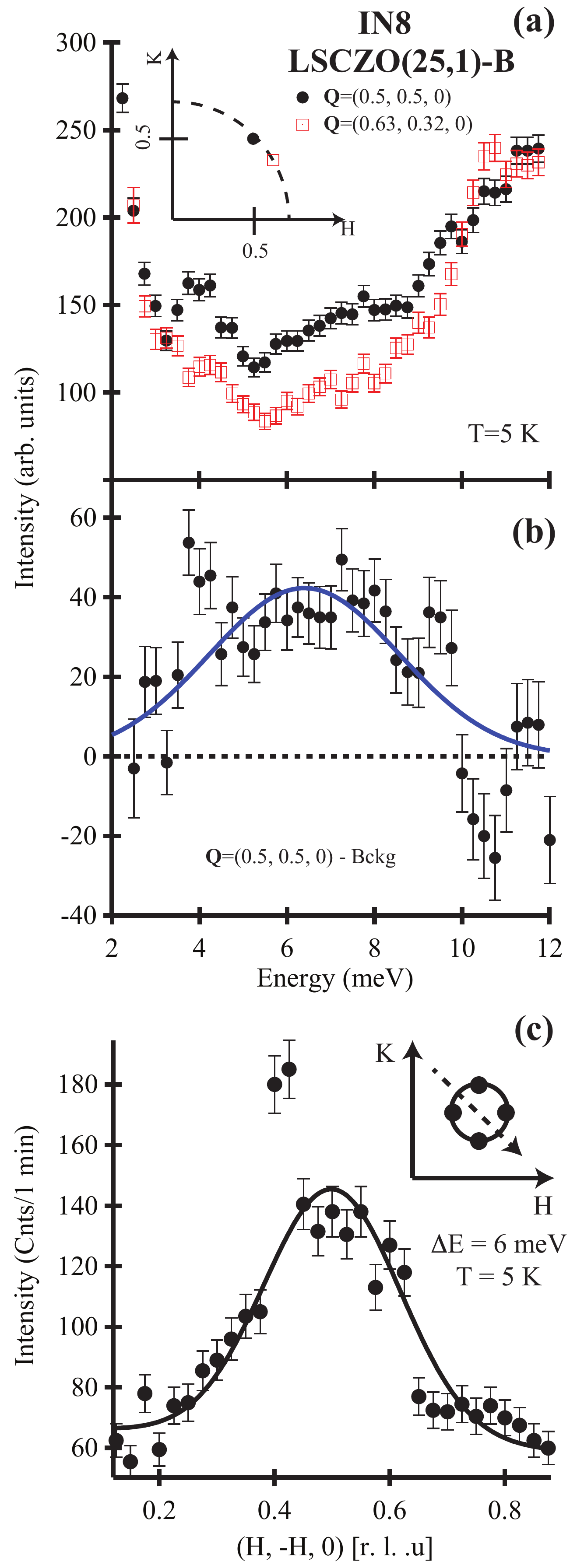}
\caption{Spin fluctuations centered at the (0.5, 0.5, 0) AF wave vector in the LSCZO(25,1)-B sample.  (a)  Raw constant Q scans centered at (0.5, 0.5, 0) and background (0.63, 0.32, 0) positions collected at T$=5$K.  Inset shows the relative signal and background positions in reciprocal space.  (b) Background subtracted spectra from panel (a) where symbols are (0.5, 0.5, 0) - (0.63, 0.32, 0) data.  Solid line is the result of a simple Gaussian fit to the resulting mode centered at $E=6.4\pm0.2$ meV. (c) Q-scan taken at 6 meV through (0.5, 0.5, 0) taken along the [H, -H, 0] direction (direction shown in inset)}
\end{figure}  

\begin{figure}
\includegraphics[scale=.35]{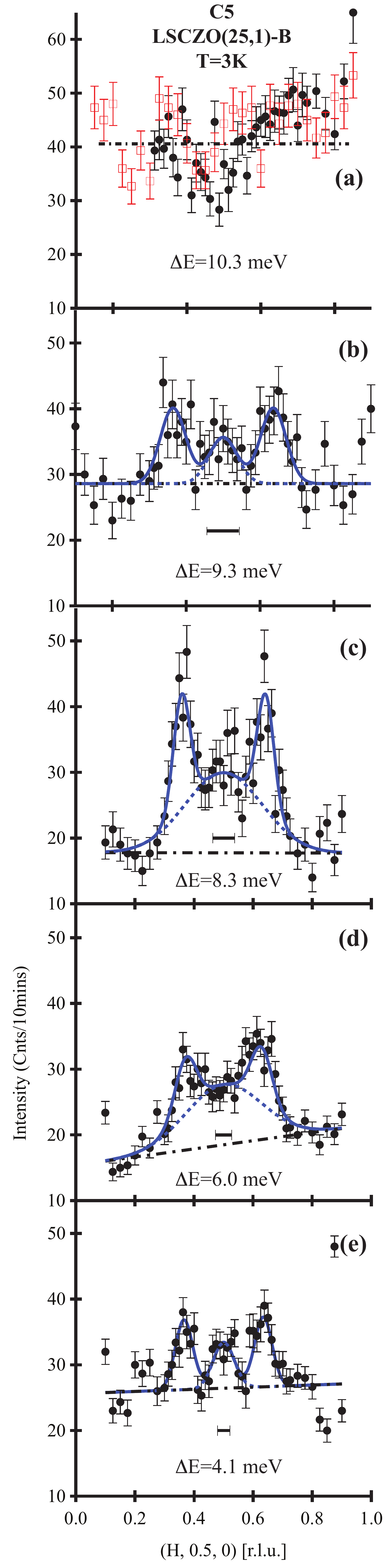}
\caption{Constant energy scans collected at T$=3$ K in the LSCZO(25,1)-B sample at various energies showing a central line-shape component centered at (0.5, 0.5, 0) in addition to the IC peak positions at (0.5$\pm\delta$, 0.5, 0), (0.5, 0.5$\pm\delta$, 0).  Solid lines are the results of symmetric Gaussian fits to IC peak positions plus an additional Gaussian peak centered at the commensurate AF position.  Dashed lines show the central Gaussian component on top of the nonmagnetic background (dash-dot lines).  Solid brackets show the momentum resolution FWHM at (0.5, 0.5, 0) at each energy.  Open squares in panel (a) are the results of a scan along K while the solid circles are scans along H.}
\end{figure}

\section{Dispersion of spin fluctuations in overdoped LSCZO(25,y)}
In order to explore the anomalous enhancement in the local spin susceptibility reported by Wakimoto et al.\cite{wakimotoLSCZO} as well as any potential changes in the effective exchange couplings coincident with this enhancement, we first measured the spin excitations of the LSCZO(25,1)-B sample from $0\leq E\leq15$ meV.  As reported previously, broadened IC spin fluctuations appeared and persisted to energies approaching $E\approx10$ meV.  Above $11$ meV however, spin fluctuations abruptly dampen beyond the resolution of our measurements (see Appendix A for further discussion).  Spin excitations were only weakly temperature dependent and showed no direct coupling to the disappearance of T$_c$ upon warming.  Explicitly, intensity spectra collected at $T=2$ K and $T=30$ K were identical within the statistics of our measurements. The resulting local susceptibility obtained by integrating the spin fluctuations across the first Brillouin zone (dashed line in inset of Fig. 1 (a)) is plotted in Fig. 1 (a).  Consistent with Ref. 5, $\chi_{2D}^{\prime\prime}(\omega)$ peaks at an enhanced value of $\chi_{2D}^{\prime\prime}(\omega)\approx 8$ $\mu_B^2$ e.V.$^{-1}$ f.u.$^{-1}$ and implies a nearly threefold enhancement in peak $\chi_{2D}^{\prime\prime}(\omega)$ in LSCZO(25,1) over LSCO(25,0) using the known $\chi^{\prime\prime}$ of Zn-free LSCO(25,0)\cite{wakimotoLSCO} (solid black line in Fig. 1 (a)).

The local spin susceptibility was also measured in a second LSCZO(25,1) sample, labeled LSCZO(25,1)-A, and is plotted in Fig. 1 (a). $\chi_{2D}^{\prime\prime}(\omega)$ in this second sample appears nearly identical to that observed in LSCZO(25,1)-B within the experimental error of normalizing between experiments on different spectrometers (discussed in Section II of this paper).  The observation of a strong enhancement in $\chi_{2D}^{\prime\prime}(\omega)$ in both samples demonstrates the robustness of this phenomenon between distinct samples.  As a reference for the expected spin response in a prototypical, Zn-free, LSCO system doped near optimal superconductivity , in Fig. 1(a) we also plot $\chi_{2D}^{\prime\prime}(\omega)$ measured for LSCO(x=0.142) under the same experimental conditions as LSCZO(25,1).  Below 4 meV, the known spin-gap in this system\cite{kofu, chang} cuts out the low energy spectral weight and $\chi_{2D}^{\prime\prime}(\omega)$ builds toward the maximum at 20 meV previously reported in LSCO(x=0.16)\cite{hayden} with low energy excitations found along the SDW order wave vector of $\delta=0.111$.  Since this sample is known to be nearly dispersionless in this energy regime, it serves as a useful reference for our experiments' ability to resolve the degree to which the dispersion is modified in the far overdoped LSCZO(25,1) samples \cite{wakimotoLSCZO, lipscombe}.  

The dispersion relations of the SDW excitations in each sample are plotted in Fig. 1 (b).  Curiously, there appears a clear dispersion in the incommensurate excitations spectrum of both LSCZO samples measured. For LSCZO(25,1)-B, the excitations seemingly disperse inward with increasing energy until they reach a minimum $\delta$ at $E=7$ meV energy transfer with $\delta_{min}=0.103\pm0.006$ r.l.u..  Upon further increase in energy, the IC excitations disperse back outward toward the zone boundary before they vanish above $\delta=0.147\pm0.009$ r.l.u. and E$=10$ meV. A similar pattern of dispersion is observed in the LSCZO(25,1)-A sample with an enhanced outward dispersion above $6$ meV where the IC excitations reach $\delta=0.179\pm0.010$ r.l.u. at $E=10$ meV.  As will be discussed in Section IV, this pattern of dispersion is a combination of two effects:  (1) unaccounted, Zn-induced, spectral weight at the (0.5, 0.5, 0) position biasing fit parameters inward, and (2) a subtle, intrinsic hour-glass-like dispersion inherent to the Zn-free LSCO(x=0.25) compound.

\section{Zn-induced Impurity Mode at \textbf{Q}$=(0.5, 0.5, 0)$}
Constituting one source of the dramatic inward dispersion observed in these Zn-doped LSCZO(25,1) samples, individual Q-scans show signs of spectral weight centered away from the IC peak positions.  Representative Q-scans at select energy transfers from experiments on both LSCZO(25,1)-A and LSCZO(25,1)-B are plotted in Fig. 2 and suggest the presence of an unaccounted line-shape component centered at the commensurate AF \textbf{Q}=(0.5, 0.5, 0) position. Further demonstrating this, energy scans taken at the \textbf{Q}$=(0.5, 0.5, 0)$ position in LSCZO(25,1)-B are overplotted with a nonmagnetic background reference scan taken at \textbf{Q}$=(0.63, 0.32, 0)$ in Fig. 3 (a).  A clear difference between the signal and background positions appears at energies above $E\approx 4$ meV and persists until $E=10$ meV.  Fig. 3 (b) shows a direct subtraction of these signal and background scans and reveals a peak in the additional spectral weight centered at $6.4$ meV; suggestive of the emergence of a localized Zn-induced impurity mode in this sample. 

Equally important, Fig. 3 (c) shows an additional Q-scan taken at $\Delta E=6$ meV through \textbf{Q}=(0.5, 0.5, 0) along the [H, -H, 0] direction that directly demonstrates the presence of a broad peak in the spectral weight centered at the AF position.  The experimental resolution of these initial measurements coupled with the broad intrinsic peak widths at the incommensurate peak positions however render a systematic separation of the two line-shape components---commensurate and incommensurate---difficult.  A similar difficulty was reported in the earlier work of Wakimoto et al. in the analysis of the enhanced $\chi_{2D}^{\prime\prime}(\omega)$ in LSCZO.  In this earlier study, the authors were unable to resolve whether additional spectral weight appeared at the commensurate (0.5, 0.5, 0) position or whether only the correlated incommensurate spin response was enhanced\cite{wakimotoLSCZO}.

In order to further explore the potential appearance of a central AF impurity resonance in overdoped LSCZO, we remeasured the spin dynamics of the LSCZO(25,1)-B sample using an instrumental setup on the C5 spectrometer with improved momentum resolution and a cleaner nonmagnetic background relative to that employed for the data collected on the HB3 and IN8 spectrometers in Figs. 1-3.  Q-scans at fixed energy transfers were repeated across the incommensurate peak positions and through the AF zone center at $3$ K, and the results of selected scans using tighter experimental collimation are plotted in Fig. 4.  From this figure, at $E=4.1$ meV, $E=8.3$ meV, and $E=9.3$ meV there appear clear signatures of commensurate AF peaks centered at (0.5, 0.5, 0) coexisting with the known IC spin excitations in this system.  As the energy is increased above $4$ meV, the inward dispersion of the IC peaks as well as a broadening of the central peak renders the three peak structure difficult to resolve due to the saddle-point in the hour-glass dispersion; however, as the IC peaks disperse back outward at energies above 7 meV, the three peak structure is again resolved.  For $E>10$ meV in Fig. 4(a), magnetic signal could no longer be reliably extracted.

\begin{figure}
\includegraphics[scale=.4]{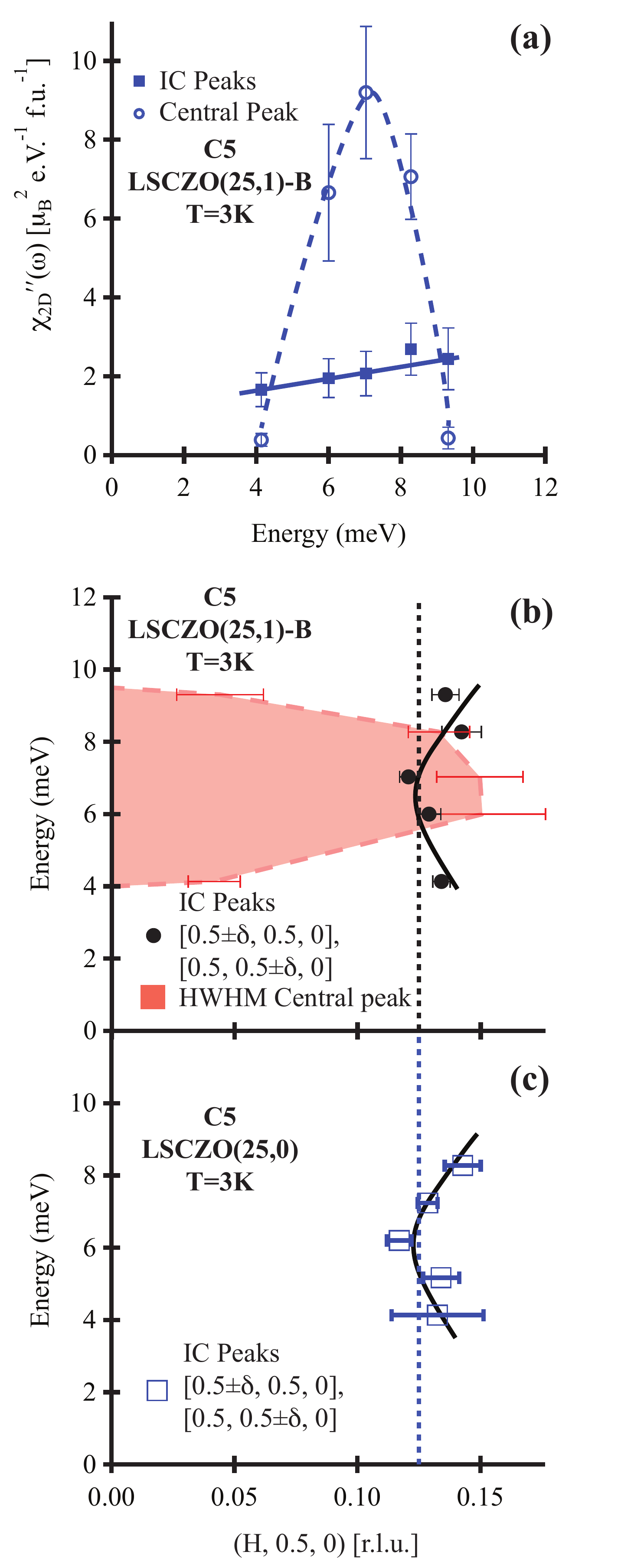}
\caption{(a) $\chi_{2D}^{\prime\prime}(\omega)$ in the LSCZO(25,1)-B sample with the contributions from the IC and commensurate line-shape components separated.  Dashed line is a guide to the eye highlighting the peak in local spin susceptibility arising from the central AF mode in this Zn-doped sample.  The solid line traces the nearly featureless $\chi_{2D}^{\prime\prime}(\omega)$ attributed to the IC excitations in this system.  (b)  Dispersion of IC spin excitations and the momentum line width of the central AF mode in LSCZO(25,1)-B.  Reanalyzing the dispersion with the central AF peak qualitatively preserves the saddle-point feature in the spectrum where the half width at half maximum (HWHM) of the commensurate peak component is largest at the minimum incommensurability.}
\end{figure}

Reanalyzing the spectral weight distribution resulting from this new picture of four IC SDW peaks coexisting with a central commensurate AF mode, the local spin susceptibility from each component is plotted in Fig. 5 (a).  An analysis identical to that employed earlier was used in integrating the spectral weight of the IC spin excitations with the addition of modeling an extra two dimensional Gaussian peak centered at (0.5, 0.5, 0).  Using this new analysis, it is found that the large gain in the spectral weight resulting from the substitution of Zn into the overdoped LSCZO(25,1) system arises almost exclusively from this new commensurate impurity mode.  The residual spectral weight arising from the IC spin excitations is similar in magnitude to that observed in pure LSCO(x=0.25)\cite{wakimotohighenergy} although the peak in $\chi_{2D}^{\prime\prime}(\omega)$ in the low energy IC spin spectrum is absent in this LSCZO(25,1)-B sample.  The inability to resolve a peaked structure in the IC spin channel of $\chi_{2D}^{\prime\prime}(\omega)$ however may be simply due to the difficulty of completely separating the spectral weight contributions from the IC and commensurate peaks at energies where the IC $\delta$ is at a minimum.  

The resulting dispersion of the IC spin excitations and their relationship to the width of the central AF excitation in this LSCZO(25,1)-B sample are plotted in Fig. 5 (b).  The width and integral area of the central peak are maximal at $E\approx7$ meV, while the hour-glass dispersion of the IC spin excitations uncovered in our earlier experiments and plotted in Fig. 1 (b) is reduced when accounting for the spectral weight at (0.5, 0.5, 0).  Excitations at the highest energy transfer and the highest incommensurability $\delta$ observed in our previous IN8 experiment (at $10$ meV) were not resolvable in this higher resolution measurement; however the outward dispersion of the IC peaks is still observed above $7$ meV.  The primary modification to the previous picture of the spin dispersion in this new dispersion plot therefore is the reduction in the overall inward dispersion of the IC spin excitations.  This reduction in the minimum incommensurability results from the removal of commensurate spectral weight that otherwise slightly biases the fit of the IC peak positions, although qualitatively the picture of a saddle point in the spin spectrum in this LSCZO(25,1)-B sample remains the same.  

In order to examine the origin of this subtle saddle point in the excitation spectrum of LSCZO(25,1), we reexamined data collected earlier on LSCO(x=0.25) under near identical experimental conditions \cite{wakimotoLSCO}.  Spin excitations from this Zn-free sample were fit using an identical model to that utilized in the current paper, and the resulting dispersion is plotted in Fig. 5 (c).  Within the error of the measurements, both the Zn-free LSCO(x=0.25) and Zn-substituted LSCZO(25,1) samples reveal a nearly identical pattern of dispersion within their incommensurate spin excitations; however the dispersion below 7 meV in LSCO(x=0.25) is within experimental error.  This suggests that the residual dispersion remaining after accounting for the AF excitations at (0.5, 0.5, 0) in LSCZO(25,1) is intrinsic to the LSCO(x=0.25) phase and not induced by Zn-substitution.        

We note here, that although an intrinsically broad momentum width of the Zn-induced commensurate AF peak is evidenced in Fig. 3 (c) and suggests that the above model of the Zn-induced fluctuations is appropriate, an alternate model of the data presenting a sharper, long-range, AF mode centered at (0.5, 0.5, 0) can also be fit to the data plotted in Fig. 4.   The results and implications of fits to this alternate model are discussed in the appendix this paper; however this alternate model is unable to account for the broad peak width known to be present at 6 meV in Fig. 3 (c).   

\section{Discussion}
The common SDW dispersion in both overdoped LSCO(x=0.25) and Zn-substituted LSCZO(25,1) samples resembles an incomplete ``hour-glass'' excitation spectrum where the minimum incommensurability occurs at the maximum in $\chi^{\prime\prime}(\omega)$---similar to peak in the local spin susceptibility of optimally doped LSCO\cite{hayden, tranquada}.  Whereas the failure of IC excitations to disperse completely inward to the commensurate AF zone center seems to be a generic feature of overdoped LSCO and has been previously observed in Zn-free LSCO(x=0.22)\cite{lipscombe}, the minimum incommensurability of LSCZO(25,1) and LSCO(x=0.25) at $E\approx 7$ meV appears at a significantly lower energy than that of the LSCO(x=0.22) system (where $\delta_{min}$ occurs at $E\approx 45$ meV). 

The energy scale at which $\delta_{min}$ occurs is widely thought to signify a crossover in dynamic spin regimes where, at energies above $\delta_{min}$, spin-wave excitations reminiscent of the high frequency spin fluctuations of the initial parent phase La$_2$CuO$_4$ appear\cite{hayden, wilsonhighenergy, stock}.  As a reflection of this, the crossover in the dispersion of the hour-glass spectrum ubiquitous in hole-doped cuprates has been suggested to scale universally with the effective superexchange coupling of the corresponding insulating parent compound\cite{tranquada}.  The dramatic reduction in the energy scale of $\delta_{min}$ upon heavily overdoping Sr may therefore imply that the residual AF exchange interactions have been substantially weakened from the parent state as the system is tuned to the far edge of the superconducting dome in its phase diagram.   

We do note however that, following the identification and removal of the central commensurate peak from the dispersion of the incommensurate spin excitations of LSCZO(25, 1), the remaining hour-glass behavior is subtle.  While our data still point toward the continued presence of an hour-glass feature with an inward dispersion at $7$ meV in LSCZO(25,1), we must also consider the possibility of a nearly nondispersive picture of the excitation spectrum below $10$ meV.  In such a scenario, the incommensurate $\delta$ would fit to an anomalously large average value of $\delta=0.132\pm0.002$---a value greater than the maximum $\delta=0.125$ thought to exist in Zn-free LSCO\cite{yamadaplot}.  Incommensurate order with $\delta>0.125$ has been previously reported in studies of magnetic Fe-impurity substitution into overdoped Bi$_{1.75}$Pb$_{0.35}$Sr$_{1.90}$CuO$_{6+z}$ \cite{hiraka} as well as into overdoped LSCO(x=0.25)\cite{he}; however the enhanced incommensurability in both cases does not apply to the incommensurate spin fluctuations, which remain centered at $\delta=0.125$.  The larger incommensurabilities manifested in the static order in Fe-doped LSCO were attributed to modified nesting wave vectors across the Fermi surfaces and reflective of a coupling between the Fe-spin and the itinerant electrons in the material\cite{he}. It is interesting to note that our data instead demonstrate that the primary effect of nonmagnetic Zn-impurities in this same overdoped LSCO(x=0.25) system is to stabilize dynamical, \textit{commensurate}, AF spin correlations, providing a strong contrast to the effect of magnetic Fe-impurities in the overdoped regime.      

Our main result however is uncovering the presence of a commensurate excitation at the AF $\textbf{Q}=(0.5, 0.5, 0)$ wave vector in Zn-doped LSCZO(25, 1).  This provides considerable insight into the origin of the anomalous enhancement in the local spin susceptibility previously reported in this same system.  The resulting localized impurity mode comprises the bulk of the spectral weight in the excitation spectrum of LSCZO(25,1) below $15$ meV where the total integrated moment of the Zn-impurity state in Fig. 5 (a) is $0.041\pm0.008$ $\mu_B^2$ compared to the $0.016\pm0.007$ $\mu_B^2$ arising from the IC spin fluctuations integrated over the same energy interval (0 to 15 meV).  No static magnetic order accompanies the emergence of this AF impurity mode confirming that the Zn-induced enhancement of AF correlations are purely a dynamic effect in the far overdoped regime of LSCO\cite{wakimotoLSCZO}.  This contrasts with previous neutron studies demonstrating an enhancement or stabilization of static SDW order in LSCZO variants with smaller Sr-content\cite{hirotareview}; however our results are consistent with $\mu$sR studies reporting a slowdown in Cu-spin fluctuations within progressively overdoped LSCZO concentrations\cite{risdiana}.      

At the peak susceptibility at 7 meV in LSCZO(25,1), the radius of the dynamic correlation length associated with the Zn-induced commensurate impurity mode can be determined by the relation $\xi=\sqrt{2ln(2)}(\frac{2\pi}{a}*w)^{-1}$ where $w=0.116$ [r.l.u.] is the Gaussian width obtained from resolution convolved fits to the data. The resulting correlation length is $\xi=6\pm1\AA$ revealing a polarized halo of AF Cu$^{2+}$ spins spanning an average diameter of three unit cells. Our experimental picture of highly local Zn-polarized magnetic impurity states, phase-separated from the SC regions, is consistent with the model of microscopic phase separation initially advanced by $\mu$sR studies\cite{nachumi}; however the mechanism of magnetic pinning in these LSCZO(25,1) samples seems to be distinct from that observed previously in optimally and underdoped LSCZO systems.  Rather than pinning stripe correlations\cite{kivelson} and static SDW order observed in more lightly Sr-doped samples\cite{kimura, hirotareview}, Zn-substitution in LSCO(x=0.25) samples instead stabilizes commensurate AF fluctuations with a momentum distribution consistent with the parent La$_2$CuO$_4$ material\cite{kastner}. This suggests the relevance of the pseudogap energy scale in the nature of Zn-induced magnetic states across the LSCO phase diagram.  LSCO(25, 1) has been doped just beyond the vanishing of the pseudogap phase at $x\approx 0.24$ \cite{taillfeur} and, simultaneous to this, the states stabilized via Zn-substitution switch from the creation of incommensurate SDW states for LSCO($x<0.24$)\cite{kimuraoverdoped} to the creation of commensurate AF fluctuations reminiscent of the parent material for LSCO(x=0.25).  
    
The downward renormalization in energy of the remnant AF spin spectrum at the expense of superconductivity suggests the continued importance of high frequency AF fluctuations remnant from the parent state within the superconducting properties of overdoped LSCZO.  This finding is in agreement with recent work studying the AF paramagnon spectra of bilayer cuprate systems demonstrating the persistence of robust, nearly unchanged, AF fluctuation spectra throughout their phase diagrams\cite{tacon}.   Our results therefore point toward a Zn-induced pair breaking mechanism rooted in the slowing down of high frequency AF fluctuations for overdoped LSCZO concentrations far away from the psuedogap regime.  This stems from our ability to directly resolve a dramatically slowed halo of polarized AF fluctuations about Zn-impurity sites that persist across a length scale consistent with their formation of extended pair breaking objects within the CuO$_{2}$ planes.  This AF fluctuation driven mechanism of pair-breaking is distinct from the suppression of SC via the stabilization of competing order such as stripe fluctuations in optimally and underdoped LSCZO and thus opens the possibility for two avenues for the Zn-induced suppression of the superconducting condensate in the cuprates.                    
    
\section{Conclusions}
Our inelastic neutron scattering measurements of LSCZO(25, 1) have resolved the emergence of a new, commensurate, AF excitation resulting from Zn-induced impurity states with a characteristic energy of $E_{impurity}=7$ meV.  This localized impurity mode comprises the entirety of the enhancement in $\chi^{\prime\prime}(\omega)$ brought on by Zn substitution into heavily overdoped LSCO systems and elucidates the origin of the Zn-induced magnetism in this overdoped material.  The appearance of this highly localized AF impurity mode upon Zn-doping and the corresponding suppression of T$_c$ suggests the importance of high frequency AF spin fluctuations within the superconducting state near the far overdoped edge of superconductivity in the cuprates.  

\acknowledgments{
SDW acknowledges helpful discussions with Ziqiang Wang.  The work at BC was supported by NSF Award DMR-1056625.  Part of this work was performed at ORNL's HFIR, sponsored by the Scientific User Facilities Division, Office of Basic Energy Sciences, U.S. Department of Energy. The work at LBNL was supported by the Director, Office of Science, Office of Basic Energy Sciences, Division of Materials Science and Engineering, of the U.S. Department of Energy under Contract No. DE-AC02-05CH11231.  The work at Tohoku University was supported by the Grant-In-Aid for Science Research A (22244039) from the MEXT of Japan.}

\section{Appendix A: Spin phonon coupling in LSCZO(25,1) }
The abrupt cutoff of the AF spin spectrum above 10 meV in LSCZO(25, 1) is accompanied by an unusual broadening within an optical phonon mode at \textbf{Q}$=(0.7, 0.5, 0)$.  This broadening is shown in Fig. 6 where scans stepping along the nuclear zone boundary (ZB) show the phonon dispersion along the $(H, 0.5, 0)$ direction.   At E$\approx13$ meV the width of Z.B. phonon peaks reaches a maximum at an energy corresponding to the projected intersection of the incommensurate spin fluctuation dispersion and the phonon energy evolution along the zone boundary. This unexpected broadening in the phonon line width suggests strongly enhanced spin-phonon coupling within this LSCZO(25, 1) system where phonon excitations with momentum \textbf{Q}$=(0.7, 0.5, 0)$ may acquire an additional decay channel at the energy where the incommensurate spin fluctuation spectrum disperses to overlap with this phonon branch.  This strong spin-phonon coupling may also account for the abrupt disappearance of the spin spectrum in this LSCZO(25,1) system through providing an enhanced dampening interaction that precludes our experiment's resolution of correlated spin excitations above $E=10$ meV.  Further systematic measurements however are required to fully explore this effect in Zn-doped LSCO systems where the dispersion of low energy spin fluctuations are dramatically modified relative to their Zn-free counterparts.

\begin{figure}[h]
\includegraphics[scale=.4]{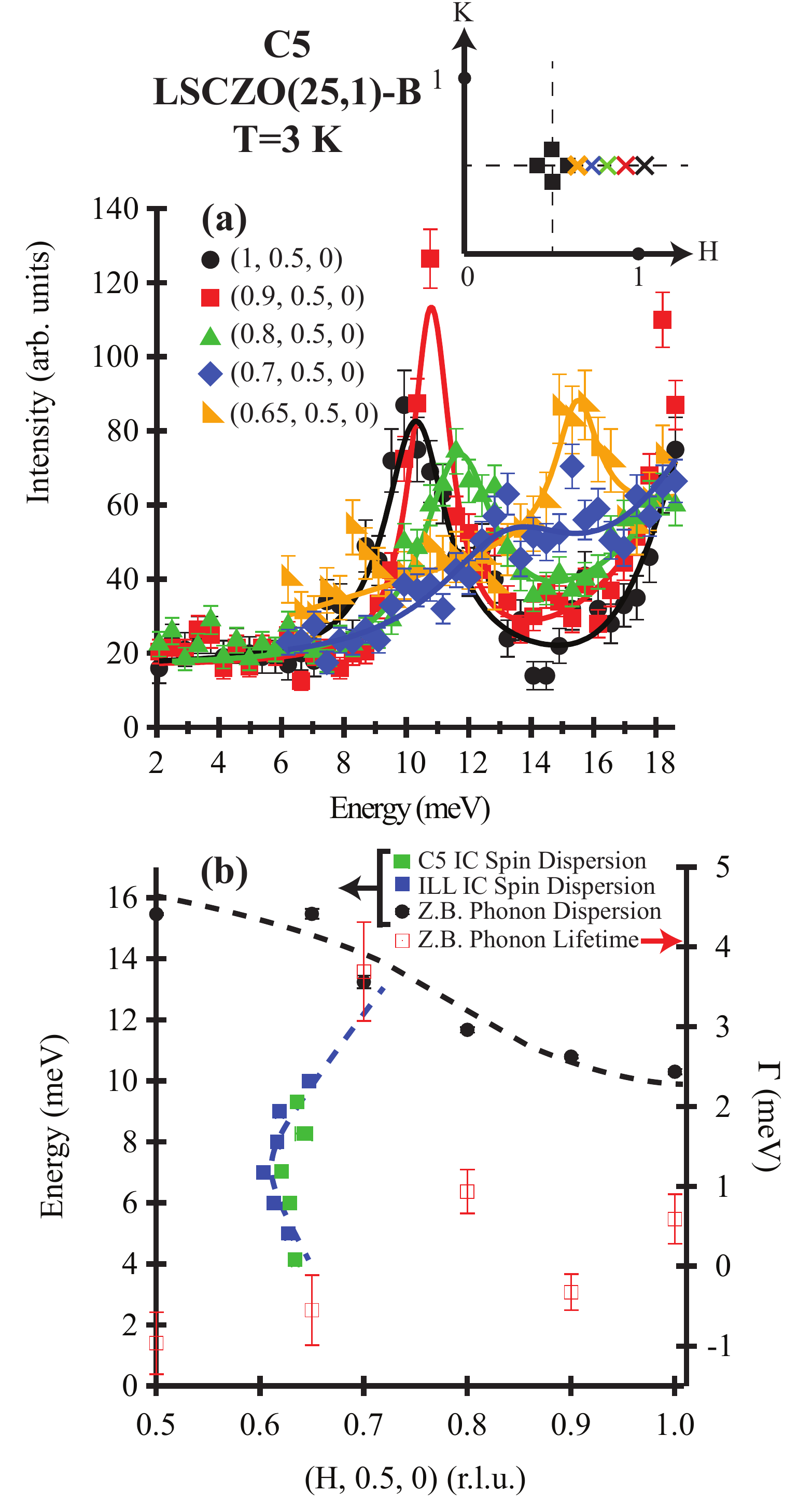}
\caption{(a)  Phonon dispersion along the chemical cell zone boundary in sample LSCZO(25,1)-B.  Constant Q scans are shown starting at (1, 0.5, 0) moving incrementally toward the expected IC magnetic peak position where the inset shows Q-positions chosen for energy scans (crosses) relative to the IC magnetic positions (solid squares).  (b)  Dispersion of resulting phonon energies obtained from Lorentzian fits to the data in panel (a) overploted with the dispersion of the IC magnetic peaks from both C5 and IN8 experiments on this same LSCZO(25, 1)-B sample.  Open squares show the deconvolved lifetime $\Gamma$ of the corresponding phonon mode as the dispersion approaches the branch of magnetic excitations.  At the projected intersection, the lifetime of the phonon mode significantly broadens and the magnetic excitation spectrum broadens beyond the experimental resolution.  Negative lifetime values result from a overly conservative deconvolution of the spectrometer resolution and should be regarded as resolution limited.}
\end{figure}

\section{Appendix B:  Alternate model of $\textbf{Q}=(0.5, 0.5, 0)$ commensurate mode}
Due to the inward dispersion of the incommensurate spin excitations near the maximum in the local spin susceptibility at $\Delta E\approx 7$ meV, there are two possible methods of fitting the incommensurate and commensurate components of the spin spectrum measured.  The first method allows the non-linear least squares routine to find a minimum in parameter space where the commensurate peak width is broad, implying a nearly localized excitation induced via Zn-impurity substitution with the resulting $\chi^{\prime\prime} (\omega)$ plotted in Fig. 5 of the main paper.  This interpretation is supported by the broad peak width observed at (0.5, 0.5, 0) in Fig. 3 (c).  If we ignore this constraint however, an alternate fit can also be performed where the momentum width of the central peak is constrained to be the width observed explicitly at $\Delta E=4.1$ meV.  

Refitting the data with this narrowly constrained central AF mode results in the alternate $\chi^{\prime\prime} (\omega)$ plotted in Fig. 7 (a).  In this scenario, the bulk of the spectral weight arises from an enhancement in the incommensurate spin excitations via Zn substitution and the appearance of the central (0.5, 0.5, 0) peak contributes minimally to the total spin susceptibility. The resulting widths of the IC and commensurate peaks from this model are plotted in Fig. 7 (b) where the IC peak widths are seen to broaden at the peak in $\chi^{\prime\prime} (\omega)$.  This alternate scenario would therefore imply two distinct effects from doping Zn into this LSCZO(25, 1) system---the appearance of a commensurate AF mode, and the enhancement of existing excitations at the IC wave vectors.

\begin{figure}[t]
\includegraphics[scale=.4]{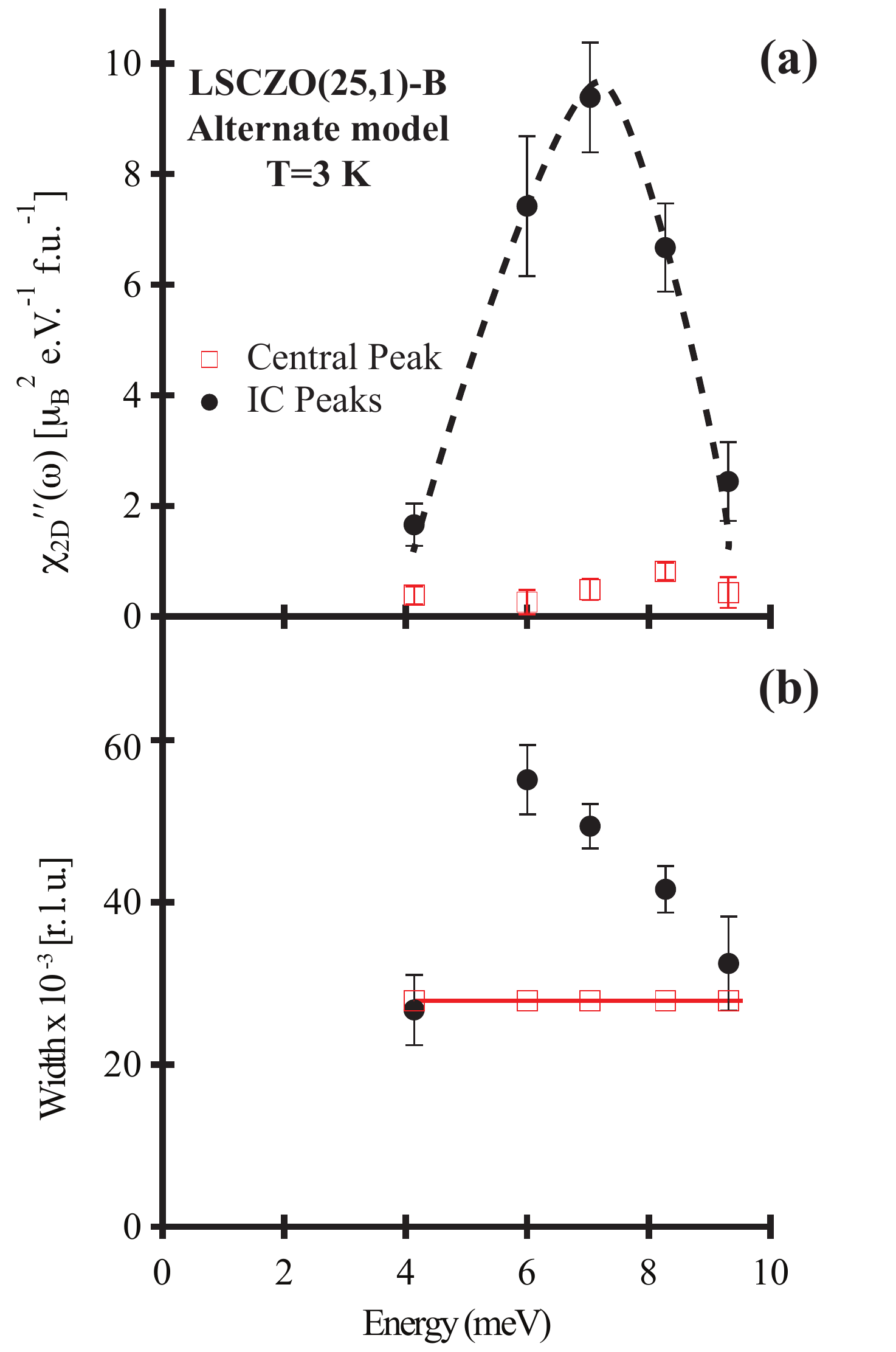}
\caption{Results of fits using an alternate model with a narrow momentum peak width at (0.5, 0.5, 0) as described in the text.  (a)  The resulting local spin susceptibility from fits using a constrained peak width at (0.5, 0.5, 0). Black circles show $\chi^{\prime\prime}(\omega)$ arising from the incommensurate peak component of the spectrum while square symbols show $\chi^{\prime\prime} (\omega)$  arising from the Zn-induced commensurate mode.  (b) Gaussian widths resulting from resolution convolved fits of both the incommensurate and commensurate peaks in Q-space.}
\end{figure}


\end{document}